\documentclass[a4paper,fleqn,usenatbib]{mnras}

\usepackage{newtxtext,newtxmath}
\usepackage{times,txfonts} 

\usepackage[T1]{fontenc}
\usepackage{ae,aecompl}

\usepackage{graphicx}	
\usepackage{amsmath}	
\usepackage{amssymb}	
\usepackage{gensymb}
\usepackage{threeparttable}
\usepackage{tablefootnote}

\title[A Quasar Discovered at redshift 6.6]{A Quasar Discovered at redshift 6.6 from Pan-STARRS1}

\author[J-J. Tang et al.]{
Ji-Jia Tang$^{1,2}$,
Tomotsugu Goto$^{3}$,
Youichi Ohyama$^{2}$,
Wen-Ping Chen$^{4}$,
\newauthor
Fabian Walter$^{5}$,
Bram Venemans$^{5}$,
Kenneth C. Chambers$^{6}$,
\newauthor
Eduardo Ba{\~n}ados$^{7,\dagger}$,
Roberto Decarli$^{5}$,
Xiaohui Fan$^{8}$,
Emanuele Farina$^{5}$,
\newauthor
Chiara Mazzucchelli$^{5}$,
Nick Kaiser$^{6}$,
Eugene A. Magnier$^{6}$,
and PS1 collaboration et al.
\\
$^{1}$Graduate Institute of Astrophysics, National Taiwan University, No.1 Sec.4 Roosevelt Rd., Taipei 10617, Taiwan\\
$^{2}$Institute of Astronomy and Astrophysics, Academia Sinica, No.1, Sec. 4, Roosevelt Rd., Taipei 10617, Taiwan\\
$^{3}$Institute of Astronomy, National Tsing Hua University, 101, Sec. 2, Kuang Fu Rd., Hsinchu 30013, Taiwan\\
$^{4}$Graduate Institute of Astronomy, National Central University, No. 300, Zhongda Rd., Zhongli Dist., Taoyuan City 32001, Taiwan\\
$^{5}$Max-Plank Institute for Astronomy, K{\"o}nigstuhl 17, D-69117 Heidelberg, Germany\\
$^{6}$Institute for Astronomy, University of Hawaii at Manoa, Honolulu, HI 96822, USA\\
$^{7}$Carnegie Observatories, 813 Santa Barbara Street, Pasadena, California, 91101 USA\\
$^{8}$Steward Observatory, University of Arizona, Tucson, AZ 85721, USA\\
$\dagger$Carnegie-Princeton Fellow
}

\date{Accepted XXX. Received YYY; in original form ZZZ}

\pubyear{2017}

\begin{document}
\label{firstpage}
\pagerange{\pageref{firstpage}--\pageref{lastpage}}
\maketitle

\begin{abstract}
Luminous high-redshift quasars can be used to probe of the intergalactic medium (IGM) in the early universe because their UV light is absorbed by the neutral hydrogen along the line of sight. 
They help us to measure the neutral hydrogen fraction of the high-z universe, shedding light on the end of reionization epoch. 
In this paper, we present a discovery of a new quasar (PSO J006.1240+39.2219) at redshift $z=6.61\pm0.02$ from Panoramic Survey Telescope \& Rapid Response System 1. 
Including this quasar, there are nine quasars above $z>6.5$ up to date.
The estimated continuum brightness is $M_\text{1450}$=$-25.96\pm0.08$. 
PSO J006.1240+39.2219 has a strong Ly~$\alpha$ emission compared with typical low-redshift quasars, but the measured near-zone region size is  $R_\text{NZ}=3.2\pm1.1$ proper megaparsecs, which is consistent with other quasars at z$\sim$6.
\end{abstract}

\begin{keywords}
quasar --- cosmic reionization
\end{keywords}



\section{Introduction}
\label{sec:I} 

Quasars or quasi-stellar objects (QSOs) are supermassive black holes (SMBHs) with accretion disks in the center of a galaxy. 
They are amongst the most luminous observable objects after the epoch of recombination at $z \sim$ 1100. High-redshift ($z\geq$ 6.0) quasars are, thus, a powerful tool to probe the early universe. 
A quasar's spectrum can be used to estimate the mass of SMBH  \citep[e.g.][]{2014ApJ...790..145D,2011Natur.474..616M,2012AJ....143..142M,2015ApJ...801L..11V}, which constrains the evolution and formation model of SMBH within a timescale of $<$1 Gyr \citep{2007ApJ...665..187L,2009ApJ...696.1798T}. 
The Gunn-Peterson (GP) \citep{1965ApJ...142.1633G} troughs in the spectrum can constrain the neutral hydrogen (\ion{H}{I}) fraction in the early universe \citep{2001AJ....122.2850B,2002AJ....123.1247F,2006NewAR..50..665F,2006MNRAS.371..769G,2011MNRAS.416L..70B,2011MNRAS.415L...1G}. 
The GP troughs observed in the $z\sim6$ quasars discovered by Sloan Digital Sky Survey (SDSS) \citep{2006AJ....132..117F,2006ApJ...652..157M} suggest that the reionization epoch of the universe ends around $z\sim6$.

High-redshift quasars can be found by a red color, between two adjacent broad bands caused by the strong intergalactic medium (IGM) absorption on the blue side of the redshifted Ly~$\alpha$ emission (1216\AA~at rest frame). 
The number of quasars we can find is limited by the survey area and depth.
After decades of searching, more than 100 quasars are found between $5.7 < z < 6.5$ from various kind of surveys \citep[e.g.][]{2016arXiv160803279B,2016ASSL..423..187M}. 
Most of them are $i$-dropouts which are very red in $i-z$ color. 
To search for quasar above $z>6.5$, $z$-dropouts are needed because the Ly~$\alpha$ line is redshifted to wavelength $\lambda\geq9000$\AA.  
Several surveys covering wavelength $\sim$ 1 $\mu$m have been dedicated to search for them and only eight are found in previous work. 
The highest redshift quasar at $z=7.085$ was found by \citet{2011Natur.474..616M} using UK infrared Telescope Infrared Deep Sky Survey (UKIDSS; \citet{2007MNRAS.379.1599L}).  
\citet{2013ApJ...779...24V} discovered three quasars above $z>6.5$ using Visible and Infrared Survey Telescope for Astronomy (VISTA) Kilo-degree Infrared Galaxy (VIKING) survey while  \citet{2015ApJ...801L..11V} found another three between $6.5<z<6.7$ using the Panoramic Survey Telescope \& Rapid Response System 1 (Pan-STARRS1 or PS1;  \citet{2002SPIE.4836..154K,2010SPIE.7733E..0EK}). 
The Subaru High-z Exploration of Low-Luminosity Quasars (SHELLQs) survey also discovered one quasar/galaxy at $z\sim6.8$ \citep{2016ApJ...828...26M}
However, the redshift of this QSO is uncertain due to the absence of strong emission lines.

In this paper, we report a discovery of a new quasar at $z=6.6$ select from PS1 with a spectroscopic confirmation. 
This is the seventh highest quasar among nine $z$-dropout quasar ($z>6.5$) known to date. 

We adopt a cosmology with $\Omega_\text{M}=0.28$, $\Omega_\text{b}=0.045$, $\Omega_\Lambda= 0.72$ and $H_\text{0}=70$ km s$^{-1}$ Mpc$^{-1}$ \citep{2011ApJS..192...18K}. 
All magnitudes are given in the AB system.

\section{Candidate Selection}
\label{sec:CS} 

PS1 covers more than $30,000$ $deg^2$ with five imaging filters ($g_\text{P1}$,$r_\text{P1}$,$i_\text{P1}$,$z_\text{P1}$,$y_\text{P1}$; \citet{2010ApJS..191..376S,2012ApJ...750...99T}).
We have selected our QSO targets from PS1 (version PV2) as follows. 
At $z > 6.5$, QSOs are the $z$-dropouts. 
As the Ly~$\alpha$ line moves to the $y$ band, thus QSOs have red $z-y$ color and are very faint or undetectable in the $z$ and bluer bands. 
At first, we selected point sources as objects with a small difference between Kron and PSF magnitudes ($-0.3 < y_\text{kron}-y_\text{psf} < 0.3$)  from the PS1 multiepoch stack data.
Further we require a very red color of $z-y >2.0$, and signal to noise ratio (SNR)  larger than 10 in $y$ band. 
Similar criteria were used by \citet{2015ApJ...801L..11V} who recently reported a discovery of 3 new $z > 6.5$ QSOs from PS1. 
We also require our candidates to have $>$24 mag in the $i$ and $z$ bands that correspond to the brightness of the sky background in PS1. 
This criterion is to select candidates below detection limits of PS1 in the $i$ and $z$ bands. To exclude spurious dropouts as a result of transient or moving single-epoch objects, we require that all of our candidates are observed and detected at least on two different dates. This ensures that each of our candidates is a real object. 

Among the objects that satisfied the criteria above, we have put priority on those with good visibility at the time of the observation, brighter $y$-band magnitude, and redder $z-y$ color.
The remaining candidates selected in this way are visually examined using the single epoch PS1 images.
Many of them are rejected by eye as they are obvious spurious, or contaminated by a nearby bright star.
This eye-rejection process is not automated nor well-controlled.
As a result, our selection procedure is heterogeneous and would not be able to yield a statistically characterizable quasar sample.
It is important future work to establish a more controlled process based on a newer release of the PS1 catalog.
We further check our candidates with Wide-field Infrared Survey Explorer catalog (WISE,\citet{2010AJ....140.1868W}), but they are not detected in the ALLWISE catalog, due to the faintness of our candidates.

\begin{table}
	\caption{The basic properties of J006.1240+39.2219.}
	\label{tab:QSO}
	\begin{threeparttable}
	\centering
	\begin{tabular}{cc} 
		\hline
		 & J006.1240+39.2219 \\
		\hline
		R.A.(J2000) & $00^\text{h}24^\text{m}29.77^\text{s}$ \\
		DEC.(J2000) & 39$\degree$13$\arcmin$18.95$\arcsec$ \\
		$z_\text{P1,PSF}$ & $25.15\pm2.98$ \tnote{a} \\
		$y_\text{P1,PSF}$ & $20.08\pm0.08$ \\
		$W1_{\text{5}\sigma}$ & $>20.5$ \tnote{b} \\
		$W2_{\text{5}\sigma}$ & $>19.7$ \tnote{b} \\
		{\ion{N}{V}} redshift ($z_\text{\ion{N}{V}}$) & $6.61\pm0.02$ \\
		$M_\text{1450}$ \tnote{c} & $-25.96\pm0.08$ \\
		$M_\text{1450}$ \tnote{d} & $-25.94\pm0.08$ \\
		$R_\text{NZ}$ (Mpc) & $3.2\pm1.1$ \\
		$R_\text{NZ,corr}$ (Mpc) & $4.3\pm1.5$ \\
		\hline
	\end{tabular}
	\begin{tablenotes}
		\item[a] The value is the output from PS1 PV2 database, which is much fainter than the detection z-band limit of PS1 (21.6; \citet{2012AJ....143..142M} ). 
		Therefore, it is untrustworthy.
		\item[b] The SNR 5 sensitivity limit\tablefootnote{\url{http://wise2.ipac.caltech.edu/docs/release/allwise/expsup/sec2_3a.html}}. 
		We adopt Table 2 in the website and check the magnitude of the closest patch to our quasar, then convert it to AB magnitude.
		There is actually one ALLWISE detection 7.6 arcsec away from our quasar, which may be a combination of this quasar and the other nearby source due to the low spatial resolution of WISE. 
		\item[c] Assume $f_\lambda\propto\lambda^{-1.5}$ \citep{2003AJ....125.1649F}.
		\item[d] Assume $f_\lambda\propto\lambda^{-1.7}$ \citep{2016A&A...585A..87S}.
        \end{tablenotes}
        \end{threeparttable}
\end{table}

\section{Follow-Up Spectroscopy and Data Reduction}
\label{sec:SDR}

\begin{figure*}
	\includegraphics[width=\textwidth]{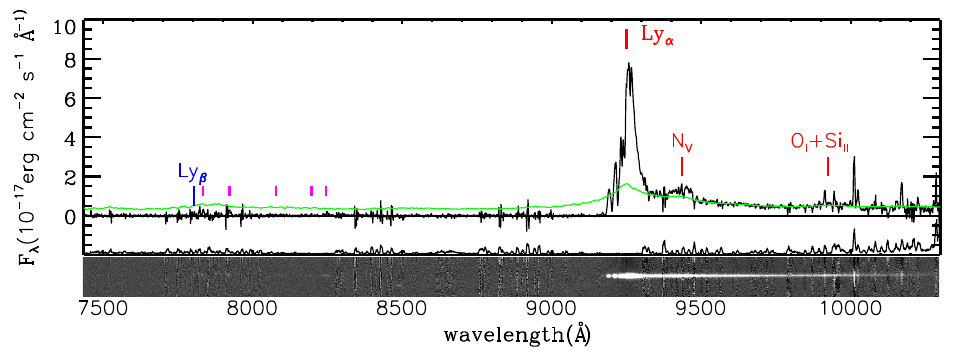}
   \caption{The spectrum of PSO J006.1240+39.2219.
   The spectrum and spectrum error are combined into upper and lower panels, respectively. 
   The spectrum error is scaled to two times larger for clarity. 
   The green line is the redshifted composite quasar spectrum from \citet{2001AJ....122..549V} scaled to the continuum flux of J006.1240+39.2219.
   Three detected lines, Ly~$\alpha$, \ion{N}{V}, and \ion{O}{I}$+$\ion{Si}{II} are marked in red. 
   Ly~$\beta$, which is marked in blue, is not detected. 
   Note that there are faint detections at 7834\AA, 7922\AA, 8078\AA, 8196\AA, and 8245\AA, which are shown in magenta.
   The two-dimensional spectrum is also shown here for comparison.}
   \label{fig:QSOspectrum}
\end{figure*}

The spectroscopic observation was carried out with the Subaru Faint Object Camera and Spectrograph (FOCAS; \citet{2002PASJ...54..819K}) on November 2, 2015. 
The weather was good and the seeing size was 0.75 arcsec. 
We used the VPH900 grating with O58 order cut filter with $2\times2$ binning, which gave us a wavelength coverage between 7434\AA~to 10582\AA~and a pixel resolution of 1.48\AA. 
We used a 0.8'' long slit, with a spectroscopic resolution of $R\sim$ 600. 
We observed 12 candidates with the first exposure of 1000 seconds.
After the first exposure, if the target did not have an emission line to be a quasar, we moved on to the next target. Most contaminating objects were late-type stars.
During the process, PSO J006.1240+39.2219 turned out to be a high-redshift quasar, noticed by a strong Ly$\alpha$ emission line.
Therefore, we took five exposures with 1000 seconds each for this target. 
Each exposure was taken at a different position on the chip to correct for systematic effects. 
We also observed BD+28D4211 as a standard star for 30 seconds.

For the data reduction, we follow the standard IRAF routines. 
We begin with flat fielding. 
The dome flat was taken for 2 seconds, and we correct for vignetting and the absorption due to the coating inside the dome. 
After normalized for the corrected flat fielding, we perform a wavelength calibration. 
We identify sky lines for 5 exposures independently and fit them into the same wavelength range (7434\AA -- 10582\AA) so that we can combine them without further wavelength transformation between different exposures later. 
We perform first order sky background subtraction using only 60 pixels in total around the quasar position in the spatial direction.
After trials and errors, this gives us the best subtraction result. 
After sky background subtraction for five exposures, we combine them using 3-$\sigma$ robust mean. 
We use \textit{apall} task in IRAF to derive 1-D spectrum from 2-D images. 
However, we need the trace from the standard star because the target is very faint bluewards of Ly~$\alpha$ emission. 
We apply the same flat fielding and wavelength calibration to the standard star, trace the spectrum using \textit{apall}, and apply it for the quasar. 
Thus, we have an accurate trace bluer than redshifted Ly~$\alpha$ emission, where we can measure the absorption.

To convert counts into flux, we use the sensitivity function derived from the standard star to correct for the wavelength dependent feature.
Then, we scale the spectrum flux to the extinction-corrected ($A_\text{y}=0.08$) PS1magnitude of $y=20.00$, which is 70\% of the spectrum flux we calibrate above.
By fitting the point spread function on the spatial direction of the 2-D image, we estimate that the slitloss of standard star to that of the target is 80\%, which is comparable to the 70\% we scale above.
The spectrum error is derived from the count dispersion between five exposures by assuming the pixel performance is the same for different exposures. 
We remove those pixels whose value is more than 3 sigma away from the mean of five exposures to eliminate cosmic ray.
Then divide the value by the square root of the number of frames left.
After considering error propagation, we turn the error into flux unit. 
This flux error is used for lines and continuum fitting, and transmission calculation in Section~\ref{sec:DAR}. 
The resulting spectrum and the 1-$\sigma$ spectrum error is presented in Figure~\ref{fig:QSOspectrum}.

\section{Data Analysis and Results}
\label{sec:DAR} 

In our spectrum, three emission lines, Ly~$\alpha$, \ion{N}{V}, and \ion{O}{I}$+$\ion{Si}{II}, are detected. 
The equivalent widths (EWs) are calculated using the wavelength ranges in \citet{2001AJ....122..549V}. 
We also integrate the same wavelength range to obtain the luminosity and the flux after subtracting the power-law continuum. 
The fitting procedures for Ly~$\alpha$, \ion{N}{V}, \ion{O}{I}$+$\ion{Si}{II}, and the continuum are described in Section~\ref{sec:propr}, Section~\ref{sec:redshift}, Section~\ref{sec:m1450}, and Section~\ref{sec:m1450}, respectively. 
We estimate the luminosities of these emission lines in Table~\ref{tab:emiline}. 

\subsection{Redshift}
\label{sec:redshift}
To measure the redshift, since Ly~$\alpha$ may be partially absorbed, we use \ion{N}{V} line (hereafter $z_\text{\ion{N}{V}}$), which is a doublet line at $\lambda$ 1238.82\AA~and 1242.81\AA. 
By using a least-squares fitting to fit a double Gaussian with same width plus linear continuum, we derive the redshift of PSO J006.1240+39.2219, which is $6.61\pm0.02$ (Table~\ref{tab:QSO}). 
We find a linear continuum is good enough because the fitting wavelength range is narrow.
The actual redshift may be influenced by the broad line property of $\ion{N}{V}$, which causes about 0.02 of redshift error. 
In Figure~\ref{fig:QSOspectrum}, we overplot the redshifted composite quasar spectrum  from \citet{2001AJ....122..549V}. The spectral resolution is adjusted to match for a fair comparison.
This shows that Ly~$\alpha$ emission of PSO J006.1240+39.2219 is much stronger than those quasars at low redshifts. 
This is the seventh highest redshift quasar known to date (Section~\ref{sec:I}).

\begin{figure}
	\includegraphics[width=\columnwidth]{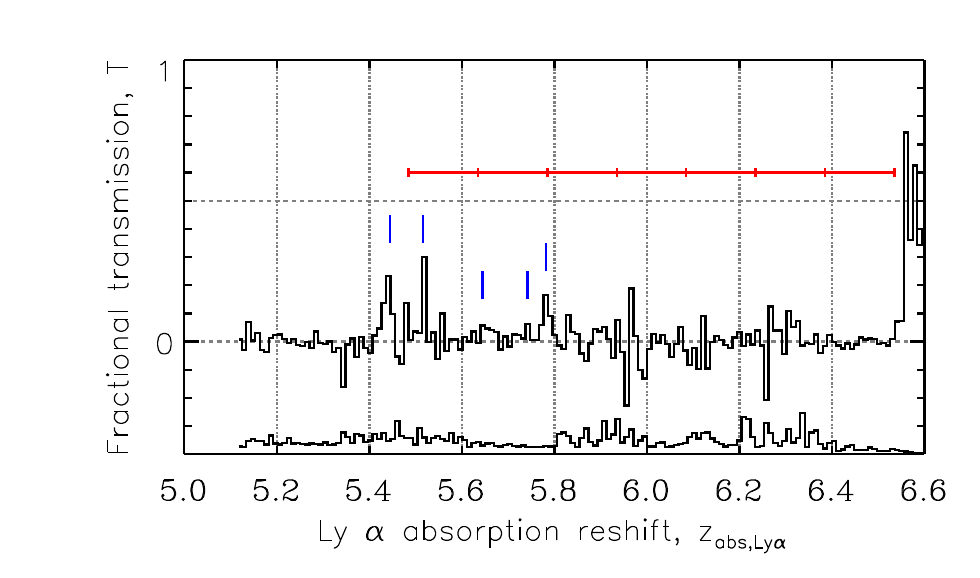}
   \caption{The T-$z$ diagram.
  This shows the transmission and the transmission error below redshifted Ly~$\alpha$ emission in terms of the Ly~$\alpha$ absorption redshift. 
   They are binned by a factor of 8 pixels for clarification. 
   There are five Ly~$\alpha$ absorption residuals labeled in the blue short vertical line. 
   The spectra shown is used to derive the effective optical depth of GP effect.
   The red horizontal line with nodes shows the bins of optical depth used in Table~\ref{tab:tau}.}
   \label{fig:zt}
\end{figure}

\subsection{$M_\text{1450}$}
\label{sec:m1450}
Conventionally, previous work used the absolute magnitude at rest frame 1450\AA ($M_\text{1450}$) to represent the quasar continuum \citep[e.g.][]{2011Natur.474..616M,2014AJ....148...14B,2015ApJ...801L..11V}. 
The $M_\text{1450}$ of PSO J006.1240+39.2219 can be derived from measuring the flux at $1450\times(z+1)$\AA. 
However, because the redshifted 1450\AA~of PSO J006.1240+39.2219 is 11032.9\AA, 
we extrapolate the continuum to estimate the flux. 
We attempt two different continuum shapes, $f_\lambda\propto\lambda^{-1.5}$ \citep{2003AJ....125.1649F} and $f_\lambda\propto\lambda^{-1.7}$ \citep{2016A&A...585A..87S}, to fit the wavelength interval between 9750\AA~to 10200\AA.
We also consider the \ion{O}{I}$+$\ion{Si}{II} lines located inside this interval.
We fit the power law continuum plus a double gaussian (\ion{O}{I}$+$\ion{Si}{II}) with least-squares fitting. 
Therefore, the estimated $M_\text{1450}$ of PSO J006.1240+39.2219 are $-25.96\pm0.08$ for $-1.5$ power law and $-25.94\pm0.08$ for $-1.7$ power law (Table~\ref{tab:QSO}), which is similar to other quasars at $z>6.5$ \citep{2015ApJ...801L..11V}. The error of $M_\text{1450}$ is dominated by the $y$-band magnitude error in PS1. 

\begin{table}
	\centering
	\caption{\em{The luminosities of the emission lines of PSO J006.1240+39.2219.}}
	{}
	\label{tab:emiline}
	\begin{tabular}{lc} 
				\hline
		Line & Luminosity\\
			& (10$^{44}$ erg s$^{-1}$) \\
		\hline
		Ly~$\alpha$ & $17.0\pm0.1$ \\
		\ion{N}{V} & $4.9\pm0.1$ \\
		\ion{O}{I}$+$\ion{Si}{II} & $2.0\pm0.1$ \\
		\hline
	\end{tabular}
\end{table}

\begin{figure}
	\includegraphics[width=\columnwidth]{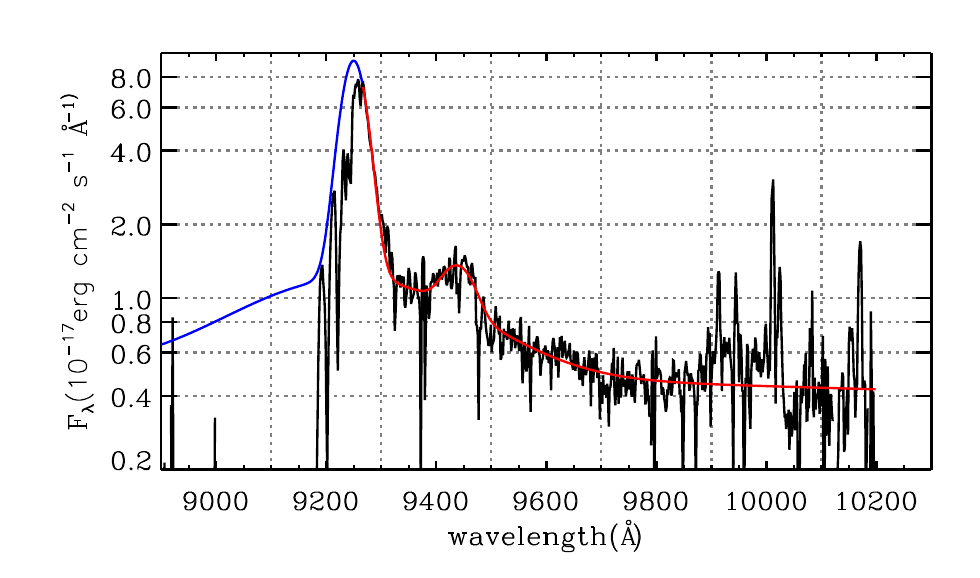}
   \caption{The simultaneous fitting of Ly~$\alpha$, \ion{N}{V}, and continuum.
   The red line is the region for fitting and the blue line is extrapolation.
   This fitting is used for the transmission calculation. 
   The fitting method is described in Section~\ref{sec:propr}.}
   \label{fig:QSOlyafit}
\end{figure}

\subsection{Optical Depth and Near Zone Region}
\label{sec:propr}
To determine GP effective optical depth $\tau_\text{GP}^\text{eff}$, we need a good estimate of the original continuum before absorption.
We cannot use the composite spectrum from low redshift quasars \citep{2001AJ....122..549V} because PSO J006.1240+39.2219 has much stronger Ly~$\alpha$ emission. 
We find a single gaussian is not suitable because there is a much broader tail of Ly~$\alpha$ emission. 
Therefore, we fit the Ly~$\alpha$ feature with a double gaussian at 1215.67\AA~\citep{2001AJ....122..549V} but with different widths with the fixed redshift at $z_\text{\ion{N}{V}}$. A similar method is used by \citet{2015MNRAS.447.2671Z}. 

\begin{figure}
	\includegraphics[width=\columnwidth]{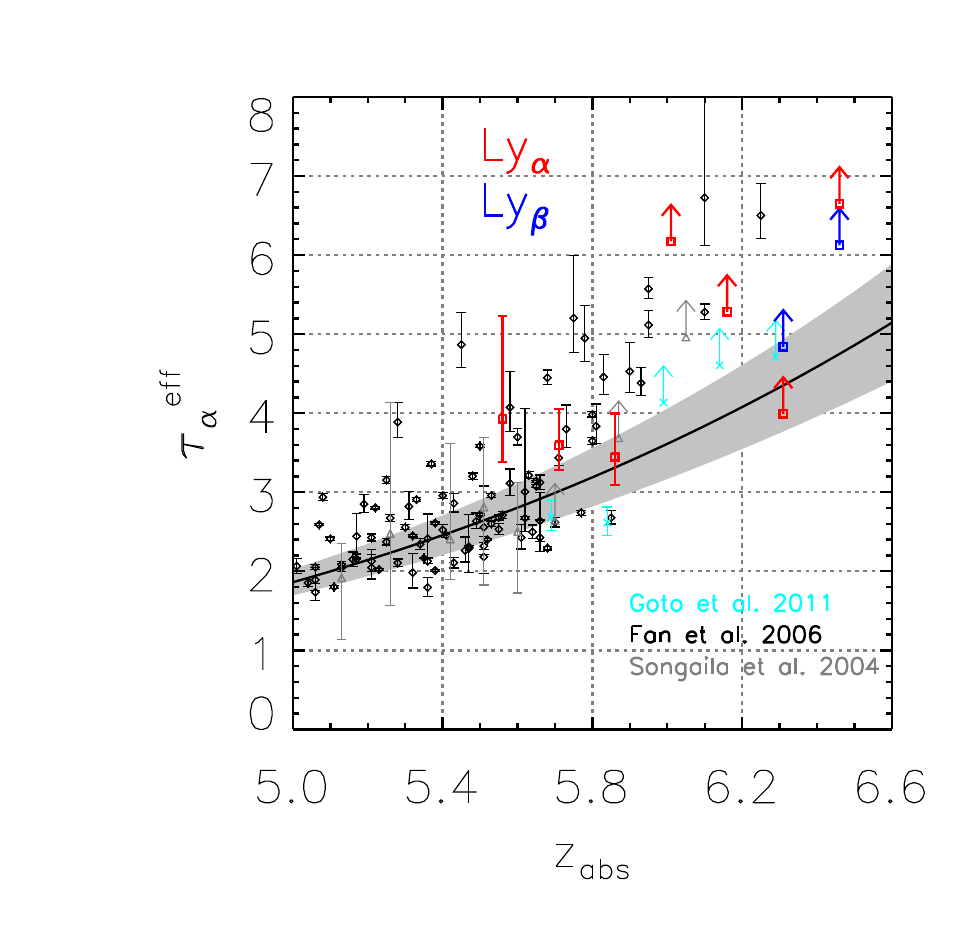}
   \caption{The GP optical depth from the spectrum of PSO J006.1240+39.2219.
   This shows the comparison of GP optical depth between PSO J006.1240+39.2219 and previous work.
   We adopt the values from Table~\ref{tab:tau} to show the optical depth from both Ly~$\alpha$ (red) and Ly~$\beta$ (blue) absorption.
   The lower limit of the GP optical depth may be lower due to the noise correlation between contiguous pixels.
   The cyan crosses with error bars show the value in \citet{2011MNRAS.415L...1G} and the 2-$\sigma$ lower limit is used for non-detection.
   The black diamonds and gray triangles with error bars are the $\tau^\text{eff}_\alpha$ in \citet{2006AJ....132..117F,2004AJ....127.2598S}, respectively.
   The solid curve with shaded region shows the best-fit power law with the error for Ly~$\alpha$ absorption at $z_\text{abs}<5.5$:$\tau^\text{eff}_\alpha=(0.85\pm0.06)[(1+z)/5]^{(4.3\pm0.3)}$ \citep{2006AJ....132..117F}.}
   \label{fig:opdz}
\end{figure}

We derive GP optical depth from the weighted transmission, $T$, in a redshift range of $5.56<z<6.46$ to avoid both Ly~$\alpha$ and Ly~$\beta$ emission.
The transmission  (Figure~\ref{fig:zt}) are measured by dividing the observed spectrum by the best-fit continuum, which includes Ly~$\alpha$ double gaussian, \ion{N}{V} double gaussian, and the continuum power-law as shown in Figure~\ref{fig:QSOlyafit}.
Then we bin the transmission within $\Delta z=0.15$ by considering the squared transmission error $\sigma_\text{T}^2$ as a variance to derive weighted average transmission.
We calculate the weighted transmission error by squared root of the reciprocal of the $\Sigma(1/\sigma_\text{T}^2)$ in the bin.
However, this error may be underestimated due to noise correlations between contiguous pixels (e.g. pixels affected by a strong sky line).
Finally, we use $\tau_\text{GP}^\text{eff}=-\ln(T)$ to calculate GP optical depth \citep{2006AJ....132..117F}.
The lower limit of the GP optical depth may be lower due to the noise correlation.
The result is shown in Table~\ref{tab:tau} and Figure~\ref{fig:opdz}.

We calculate the weighted transmission of Ly~$\beta$ in the wavelength below rest frame 1017\AA, considering, at the same time, the Ly~$\alpha$ absorption at the same wavelength based on the equation (5) in \citet{2006AJ....132..117F}.
The weighted transmission error of Ly~$\beta$ is determined by both the uncertainty of Ly~$\alpha$ absorption equation and the observed spectrum error, where the former error dominates.
After obtaining $\tau_\beta$ from the transmission, we apply the $\tau_\alpha$/$\tau_\beta=2.25$ from the discussion in \citet{2006AJ....132..117F} to derive $\tau_\alpha$ for Ly~$\beta$ optical depth.
The results are shown in Table~\ref{tab:tau} and Figure~\ref{fig:opdz}.

\begin{figure}
	\includegraphics[width=\columnwidth]{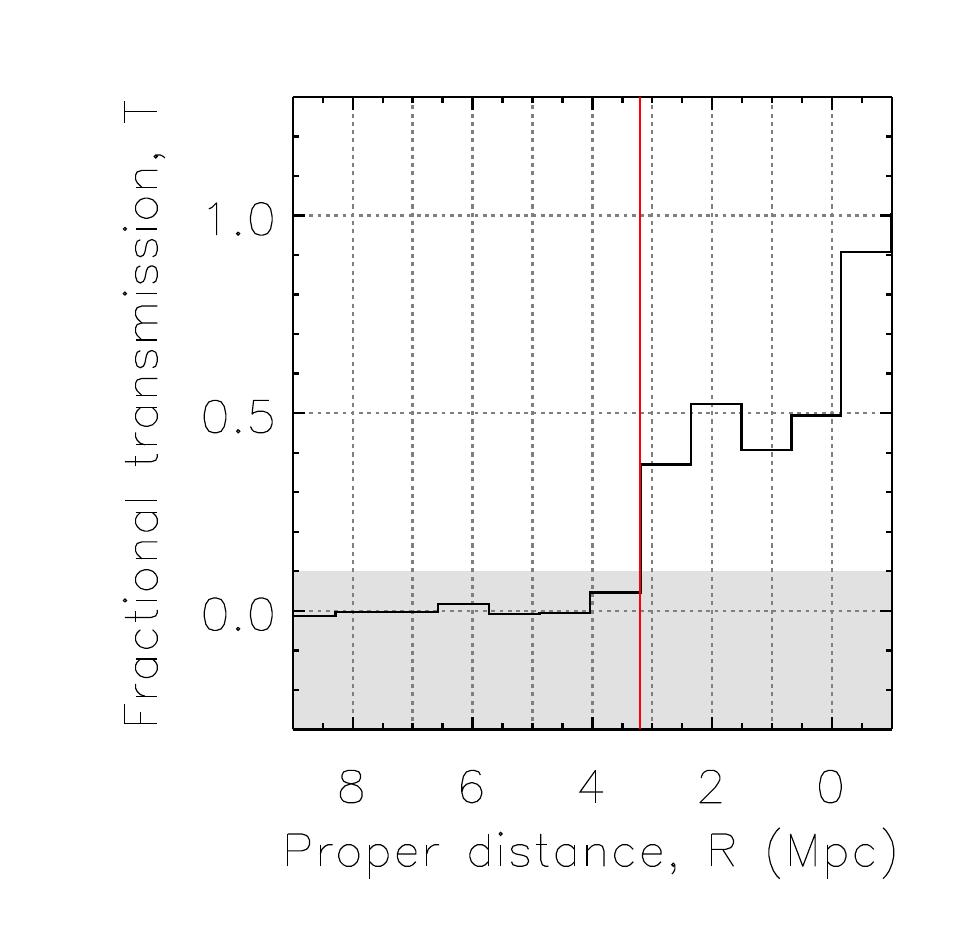}
   \caption{The T-$R_\text{p}$ diagram.
   This shows the transmission, which is smoothed for 20\AA , below redshifted Ly~$\alpha$ emission in terms of the proper distance $R_\text{p}$. 
   The red line shows $R_\text{NZ}$. 
   The error of $R_\text{NZ}$ is estimated by the uncertainty of redshift $z_\text{\ion{N}{V}}$.}
   \label{fig:rpt}
\end{figure}

We follow the following equation of \citet{2010ApJ...714..834C} to calculate near-zone region size $R_\text{NZ}$. 
\begin{equation} 
	R_\text{p,NZ}=(D_\text{Q} - D_\text{GP})/(1+z_\text{Q})
	\label{eq:rnz}
\end{equation}
The $D_\text{Q}$ and $D_\text{GP}$ are the comoving distances of the redshift of the quasar host galaxy ($z_\text{Q}$) and the redshift where the transmission drops below 0.1 ($z_\text{GP}$), respectively.
We present the proper distance $R_\text{p}$-$T$ diagram in Figure~\ref{fig:rpt}. 
The transmission is smoothed to 20\AA. 
The near-zone region $R_\text{NZ}$ is defined by the $R_\text{p}$ where $T\sim0.1$.
To reduce the effect of different luminosity, we correct the near-zone region by $R_\text{NZ, corr}=10^{0.4\times (27+M_\text{1450,AB})/3}R_\text{NZ}$ \citep{2006AJ....132..117F}. 
We measure $R_\text{NZ}=3.2\pm1.1$ (Mpc) and $R_\text{NZ,corr}=4.3\pm1.5$ (Mpc) (Table~\ref{tab:QSO}).
The error of $R_\text{NZ}$ are calculated from the error propagation of Equation~\ref{eq:rnz} assuming $\Delta_{z_\text{Q}}=0.02$ and $\Delta_{z_\text{GP}}=0.01$ \citep{2010ApJ...714..834C}.
The near-zone region of PSO J006.1240+39.2219 is consistent with other comparably distant quasars, although it has much stronger Ly~$\alpha$.

\begin{table}
	\caption{\em{The transmission and the GP optical depth}}
	{This shows the weighted transmission and effective optical depth of binned redshift interval for Ly~$\alpha$ and Ly~$\beta$. 
	Both transmission and optical depth error are in 2-$\sigma$ error if not specified.
	Those three bins in Ly~$\alpha$ which show positive transmission after considering errors are caused by the faint detections in Figure~\ref{fig:QSOspectrum}.}	\begin{threeparttable}
	\centering
	\label{tab:tau}
	\begin{tabular}{cccc} 
		\hline
		redshift & Line & Transmission & $\tau_\alpha$ \\
		\hline
		6.46 & Ly~$\alpha$ & $-0.004\pm0.006$ & $>6.6$ \\
		6.31 & Ly~$\alpha$ & $0.007\pm0.011$ & $>4.0$ \\
		6.16 & Ly~$\alpha$ & $-0.010\pm0.016$ & $>5.3$ \\
		6.01 & Ly~$\alpha$ & $-0.022\pm0.012$ & $>6.2$ \tnote{a} \\
		5.86 & Ly~$\alpha$ & $0.032\pm0.013$ & $3.4^{+0.5}_{-0.4}$ \\
		5.71 & Ly~$\alpha$ & $0.027\pm0.010$ & $3.6^{+0.5}_{-0.3}$ \\
		5.56 & Ly~$\alpha$ & $0.020\pm0.014$ & $3.9^{+1.3}_{-0.5}$ \\
		\\
		6.46 & Ly~$\beta$ & $-0.085\pm0.146$ & $>6.1$ \\
		6.31 & Ly~$\beta$ & $-0.010\pm0.120$ & $>4.8$ \\
		\hline
	\end{tabular}
	\begin{tablenotes}
		\item[a] Since the transmission is too negative, we have to use 4-$\sigma$ limit to show the optical depth.
	\end{tablenotes}
	\end{threeparttable}
\end{table}

\section{Summary}
\label{sec:last} 

We discovered a new quasar PSO J006.1240+39.2219 at $z=6.61\pm0.02$, which is the seventh highest redshift quasar known to date. 
The number of $z$-dropout ($z>6.5$) quasars reaches nine after the discovery of PSO J006.1240+39.2219. 
The rest-frame UV luminosity $M_\text{1450}=-25.96\pm0.08$ ($-25.94\pm0.08$) is comparable to other $z$-dropout quasars, but the Ly~$\alpha$ emission is much stronger than typical quasars obtained from low redshift \citep{2001AJ....122..549V}. 
The $R_\text{NZ,corr}=4.3\pm1.5$ (Mpc) of PSO J006.1240+39.2219 follows the trend $R_\text{NZ,corrected}= (7.2\pm0.2)-(6.1\pm0.7)\times(z-6)$ derived from \citet{2015ApJ...801L..11V}.

\section*{Acknowledgments}

We thank the anonymous referee for many insightful comments.
We acknowlege L.Cowie for useful discussion.
The Pan-STARRS1 Surveys (PS1) have been made possible through contributions by the Institute for Astronomy, the University of Hawaii, the Pan-STARRS Project Office, the Max-Planck Society and its participating institutes, the Max Planck Institute for Astronomy, Heidelberg and the Max Planck Institute for Extraterrestrial Physics, Garching, The Johns Hopkins University, Durham University, the University of Edinburgh, the Queen's University Belfast, the Harvard-Smithsonian Center for Astrophysics, the Las Cumbres Observatory Global Telescope Network Incorporated, the National Central University of Taiwan, the Space Telescope Science Institute, and the National Aeronautics and Space Administration under Grant No. NNX08AR22G issued through the Planetary Science Division of the NASA Science Mission Directorate, the National Science Foundation Grant No. AST-1238877, the University of Maryland, Eotvos Lorand University (ELTE), and the Los Alamos National Laboratory.
TG acknowledges the support by the Ministry of Science and Technology of Taiwan through grant
NSC 100-2112-M-001-001-MY3,
NSC 103-2112-M-007-002-MY3,
104-2112-M-001-034-,
105-2112-M-007-003-MY3,  and
105-2112-M-001-024-.
We thank Dr. Ekaterina Koptelova for the assistance in the observation.












\bsp	
\label{lastpage}
\end{document}